\documentclass[aps,prb,reprint,superscriptaddress]{revtex4-2}

\usepackage[utf8]{inputenc}
\usepackage{amsthm}
\usepackage{amsmath}
\usepackage{xfrac}
\usepackage{color}
\usepackage{hyperref}
\usepackage{graphicx}
\usepackage[dvipsnames]{xcolor}
\usepackage{amssymb}
\usepackage{bbold}
\usepackage{gensymb}
\usepackage{afterpage}
\usepackage[normalem]{ulem}

\newcommand{\ii}{\mathrm{i}} 
\newcommand{\eul}{\mathrm{e}} 
\newcommand{\id}{\mathbb{1}} 

\newcommand{\ket}[1]{|#1\rangle} 
\newcommand{\braket}[1]{\langle#1\rangle} 
\newcommand{\ketbra}[2]{|#1\rangle\!\langle #2|} 
\newcommand{\tr}{\mathrm{tr}} 

\newcommand{\rhosa}{\rho_{\Sys + \Anc}}

\newcommand{\Dyn}{\mathcal{D}}
\newcommand{\cpt}{\mathcal{E}}
\newcommand{\choi}{\chi}
\newcommand{\eoa}{F^\sharp}
\newcommand{\eof}{F}

\newcommand{\dynamics}[1]{(#1)}

\newcommand{\Sys}{\mathrm{sys}} 
\newcommand{\Anc}{\mathrm{anc}} 

\theoremstyle{plain}
\providecommand{\definitionname}{Definition}

\newcommand{\conc}{\mathcal{C}}

\newtheorem{theorem}{Theorem}

\usepackage{tikz}
\usetikzlibrary {shadows.blur, decorations.pathreplacing, arrows, positioning} 

\newcommand{\qwire}[3]{
	\draw[line width=1.2pt, blur shadow={shadow blur steps=5, shadow xshift=0.3mm, shadow yshift=-0.3mm}] (#1, #2) -- (#1+#3, #2);
}

\newcommand{\qwirev}[3]{
	\draw[line width=1.2pt, blur shadow={shadow blur steps=5, shadow xshift=0.3mm, shadow yshift=-0.3mm}] (#1, #2) -- (#1, #2+#3);
}

\newcommand{\qbox}[4]{
	\filldraw[line width=1.2pt, rounded corners, fill=GreenYellow!50!white,  draw=TealBlue!60!black, blur shadow={shadow blur steps=5, shadow xshift=0.3mm, shadow yshift=-0.3mm}] (#1, -#2-1.4) rectangle ++(#4,1.7) node[midway]{#3};
}

\newcommand{\qboxs}[4]{
	\filldraw[line width=1.2pt, rounded corners, fill=GreenYellow!50!white,  draw=TealBlue!60!black, blur shadow={shadow blur steps=5, shadow xshift=0.3mm, shadow yshift=-0.3mm}] (#1, -#2-1.4) rectangle ++(#4,1) node[midway]{#3};
}

\newcommand{\qcirc}[3]{
	\filldraw[line width=1.2pt, rounded corners, fill=BlueGreen!30!white,  draw=TealBlue!60!black, blur shadow={shadow blur steps=5, shadow xshift=0.3mm, shadow yshift=-0.3mm}] (#1, #2) circle (0.3cm) node {#3};
}

\newcommand{\wiredots}[2]{
     \qwire{#1}{#2}{0.1}
     \qwire{#1+0.2}{#2}{0.1}
     \qwire{#1+0.4}{#2}{0.1}
}

\newcommand{\initialbox}[4]{
    \qwire{#1}{#2}{-0.5}
    \qcirc{#1-0.8}{#2}{#3}
    \qwire{#1+0.7}{#2}{1.3}
    \qwire{#1-0.5}{#2-1.1}{0.5}
    \qwire{#1+0.7}{#2-1.1}{0.3}
    \qbox{#1}{0}{#4}{0.7}
}

\newcommand{\centerbox}[3]{
    \qwire{#1-0.3}{#2-1.1}{0.3}
    \qwire{#1+0.7}{#2-1.1}{0.3}
    \qbox{#1}{0}{#3}{0.7}
}

\newcommand{\wirelinedots}[2]{
    \qwire{#1}{#2}{0.3}
    \wiredots{#1+0.4}{#2}
    \qwire{#1+1}{#2}{0.3}
}

\newcommand{\finalbox}[4]{
    \qwire{#1}{0}{0.3}
    \qwire{#1}{0}{0.3}
    \qwire{#1+0.7}{0}{0.5}
    \qwire{#1+0.7}{#2-1.1}{0.5}
    \qwire{#1-0.3}{#2-1.1}{0.3}
    \qbox{#1}{0}{#4}{0.7}
    \qcirc{#1+1.5}{0}{#3}
}

\newcommand{\finalboxcustom}[5]{
    \qwire{#1}{0}{0.3}
    \qwire{#1}{0}{0.3}
    \qwire{#1+0.7}{0}{0.5}
    \qwire{#1+0.7}{#2-1.1}{0.5}
    \qwire{#1-#5}{#2-1.1}{#5}
    \qbox{#1}{0}{#4}{0.7}
    \qcirc{#1+1.5}{0}{#3}
}

\begin{document}

\title{Verifying Quantum Memory in the Dynamics of Spin Boson Models}

\author{Charlotte Bäcker}
\thanks{These authors contributed equally to this work.}
\affiliation{Institut f{\"u}r Theoretische Physik, Technische Universit{\"a}t Dresden, 
D-01062, Dresden, Germany}

\author{Valentin Link}
\thanks{These authors contributed equally to this work.}
\affiliation{Institut für Physik und Astronomie, Technische Universität Berlin, D-10623, Berlin, Germany}

\author{Walter T. Strunz}
\affiliation{Institut f{\"u}r Theoretische Physik, Technische Universit{\"a}t Dresden, 
D-01062, Dresden, Germany}

\begin{abstract}
We investigate the nature of memory effects in the non-Markovian dynamics of spin-boson models. Local quantum memory criteria can be used to indicate that the reduced dynamics of an open system necessarily requires a quantum memory in its environment. We focus on two such criteria, one based on the dynamical map and the other on single-intervention process tensors, and discuss their connection. We apply these criteria to quench dynamics in spin-boson and two-spin-boson models, computed using a numerically exact method for non-Markovian open system dynamics based on matrix product operator influence functionals. This method is ideally suited to simulate quantum multi-time processes in spin-boson models across broad parameter regimes. We find that, with access to single-intervention process tensors, one can generally predict quantum memory in the dynamics at low temperatures. Given instead only the dynamical map, it is still possible to detect quantum memory in the case of resonant environments at short evolution times. Moreover, we confirm quantum memory in the stationary dynamical regime using process tensors with the correlated steady state of system and environment as initial condition.

\end{abstract}\maketitle

\section{Introduction}

Quantum systems unavoidably interact with their environment, resulting in continuous exchanges of energy and information that introduce noise and dissipation into the dynamics. While the widely used Gorini-Kossakowski-Sudarshan-Lindblad (GKSL) master equation \cite{GorKosSud1976,Lin1976b} effectively describes memoryless quantum dynamics, it applies only to a limited subset of real-world systems. In general, in open quantum system dynamics, memory effects occur which give rise to new dynamical properties absent in the Markovian case \cite{de2017Dynamics, vacchiniOpen}. These play a crucial role in many practical applications, including quantum information processing, control, and error correction in quantum circuits \cite{reich2015Exploiting,puviani2025Non, MotMohAkh2021, OriArrFerFraBenPalFalSciMat2015}. 
Driven by these prospects, non-Markovian open quantum dynamics has emerged as an active field or research in recent decades, encompassing conceptual \cite{BreLaiPii2009, RivHuePle2010, xnpernice2012Systemenvironment, HalCreLiAnd2014, BreLaiPiiVac2016,vacchiniOpen,de2017Dynamics, PRXQuantum.2.030201, PhysRevLett.123.040401}, numerical \cite{hughes2009Effective,kolli2011Electronic,mccutcheon2011A,suess2014Hierarchy, strathearn2018Efficient, tamascelli2019Efficient,makri2020Small,makri2020Small,tanimura2020Numerically,Cygorek2021Jan,link2023Non,anto2023Effective,lacroix2024MPSDynamics, Link2024May}, and experimental approaches \cite{ApoDiPlaPat2011, LiuLiHuaLiGuoLaiBrePii2011, WhiHilPolHolMod2020}. 

Non-Markovian effects naturally arise from strong coupling to quantum reservoirs, but they can also be generated by combining Markovian quantum dynamics with non-Markovian feedback control. In fact, already a random unitary driving of the system with colored noise will typically lead to non-Markovian dynamics of the driving-averaged system state \cite{grotz2006Quantum}. {Furthermore, it is known that the probabilistic mixing of Lindblad dynamics can lead to non-Markovian dynamics \cite{MegChrPiiStr2017}.} This poses the fundamental question of whether memory effects occurring in a given non-Markovian quantum evolution can in principle be captured by classical variables  \cite{CreFac2010, vacchiniClassicalAppraisalQuantum2012,  MegChrPiiStr2017, OppSpaSodWel2023}, or if they are genuinely quantum.
Recently, local definitions and criteria of quantum memory have been put forward, which make it possible to determine whether quantum memory in the environment is strictly necessary for a given open system dynamics \cite{MilEglTarThePleSmiHue2020, Giarmatzi2021Apr,BerGarYadModPol2021, Abi2023b, BanMarHorHor2023,Backer2024Feb, TarQuiMurMil2024,backer2025Entropic}. The various approaches differ primarily by their assumptions about accessible dynamical information.  This could either be time-local information as encoded in dynamical maps, or more general multi-time statistics described by quantum process tensors. {We focus on two common criteria applicable in different frameworks, either considering a pair of quantum channels generated by the dynamical map at two subsequent times \cite{Backer2024Feb} or a process tensor with a single intervention \cite{Giarmatzi2021Apr}. We point out that, since the channels are given by marginals of the process-tensor, the map criterion is a less sensitive witness of quantum memory in the environment compared to single-intervention process tensors, but it is more easily accessible through tomography. } 

These criteria have {separately} been applied to simple toy models such as non-Markovian amplitude damping \cite{Backer2024Feb} or single spin environments \cite{Giarmatzi2021Apr}.
Such environments are highly coherent, small scale systems that, if realized experimentally, need to be engineered in quantum simulators or quantum computers. Moreover, quantum memory has recently been investigated in continuous variable systems \cite{backer2025Entropic} and spontaneous emission processes \cite{yu2025Quantum}. Non-Markovian dynamics also arises from genuine noise in real-world physical setups such as quantum dots, solid-state qubits, or impurities in condensed matter systems \cite{de2017Dynamics}. In these systems the environments consist of a continuum of degrees of freedom that can usually be assumed to obey Gaussian statistics with possibly long-time temporal correlations.  
Simulations of quantum dynamics in such environments are nontrivial, and these systems can feature rich correlated phenomena such as phase transitions \cite{vojta2006Impurity} or spin screening \cite{hewson1997The}. Nevertheless, today, efficient methods are available to simulate strongly coupled non-Markovian open quantum systems on classical computers \cite{hughes2009Effective,suess2014Hierarchy, strathearn2018Efficient, tamascelli2019Efficient,makri2020Small,makri2020Small,tanimura2020Numerically,Cygorek2021Jan,link2023Non,anto2023Effective,lacroix2024MPSDynamics, Link2024May}.

In this paper we investigate the quantum or classical nature of memory effects occurring in quantum dynamics of non-Markovian open systems with continuous Gaussian environments, focusing on spin-boson models. The numerical analysis is made possible through developments in simulation methods that use compressed representations of the bath influence functional via matrix product operators \cite{Jorgensen2019Dec,Ng2023Mar,thoenniss2024Efficient,Link2024May,sonner2025Semi}. These methods give direct access to the full multi-time dynamical propagator (process tensor) in strongly coupled regimes. We apply quantum memory criteria to the dynamics of the spin-boson model and the two-spin-boson model, considering different bath structures. {We find that it is difficult to detect signatures of quantum memory using only time-local information encoded in the dynamical map, as the corresponding criterion only predicts quantum memory when sharp resonances in the bath are present. In contrast, the process tensor yields a more sensitive detection of quantum memory present in the environment, generally indicating non-classical memory effects when the bath temperature is low.} {Process tensors also enable the study of dynamics in cases where the system is in equilibrium with the bath, such that the system-and-bath initial state is correlated. We show here that, in spin-boson models, quantum memory can be verified in this stationary regime.}

The paper is structured as follows. First, we introduce the general theoretical framework of open system dynamics with process tensors and dynamical maps in Sec.~\ref{sec:open_system_dynamics}. In Sec.~\ref{sec:influence_matrix} we describe how dynamical propagators corresponding to Gaussian environments can be computed numerically with the influence matrix approach. In Sec.~\ref{sec:qm_crit} we introduce the two common local criteria that can be used to detect quantum memory in the {environment} {and discuss how they are related. Our main results, the application to spin-boson models, are presented in Sec.~\ref{sec:spin_boson} for quench dynamics, and in Sec.~\ref{sec:stat_qm} for stationary dynamics.} Finally, we provide our conclusions and an outlook in Sec.~\ref{sec:conclusions}.

\section{Open system dynamics with process tensors}\label{sec:open_system_dynamics}

In this section we present the general framework for non-Markovian open system dynamics in terms of process tensors. A more thorough introduction to this formalism can be found in Ref.~\cite{Pollock2018Jan}. In the numerical investigation presented below, we consider specifically the spin-boson model, where a single spin is coupled to a bosonic reservoir. This model is introduced in more detail in section \ref{sec:spin_boson}. The broader theoretical framework used in this work is applicable to general quantum dynamics, so that we do not need to specify the model at this point. 

We describe the unitary dynamics of a full quantum mechanical system-and-bath model via successive unitary evolutions. In particular, the unitary channel $\mathcal{Q}$ for evolution with a small time step $\delta t$ is given as
\begin{equation}\label{eq:Q_def}
    \mathcal{Q}\rho_\mathrm{tot} = \eul^{-\ii H \delta t} \rho_\mathrm{tot}\eul^{\ii H \delta t},
\end{equation}
where $\rho_\mathrm{tot}$ is a state in the total Hilbert space of system and bath, and $H$ is the full Hamiltonian of the model. The local system state after $t_f$ time steps of length $\delta t$ can be obtained as
\begin{equation}\label{eq:unitary_evo}
    \rho_{t_f}=\tr_\mathrm{env}(\mathcal{Q})^{t_f}\rho_\mathrm{tot},
\end{equation}
where $\tr_\mathrm{env}$ denotes the trace over the environment degrees of freedom. In the process tensor formalism one allows for general ``interventions" or ``controls" on the system after a given number of time steps. Interventions can be {arbitrary local completely positive (CP) maps, such as quantum channels and measurements.}
Let $\mathcal{A}_m$ denote a set of $M+1$ interventions at time steps $t_m$ (one can set the initial time to $t_i=0$).
Then the final system state after $t_f$ time steps can be expressed as
\begin{equation}\label{eq:pt_def1}
   \begin{split}
       &\mathcal{T}_{t_f,t_M,...,t_1}[\mathcal{A}_M,...,\mathcal{A}_1,\mathcal{A}_0]=\\&\tr_\mathrm{env} (\mathcal{Q})^{t_f-t_M} \mathcal{A}_{M}(\mathcal{Q})^{t_{M}-t_{M-1}} \mathcal{A}_{M-1}\cdots(\mathcal{Q})^{t_1}\mathcal{A}_{0}\rho_\mathrm{tot}
   \end{split} 
\end{equation}
The corresponding process tensor $\mathcal{T}_{t_f,t_M,...,t_1}$ is the multi-linear map of interventions $\{\mathcal{A}_m\}$ to the final reduced system state \cite{Pollock2018Jan}.
Specifying a basis for the interventions, the process tensor can be seen as a tensor with $M+1$ indices. Evidently, this object encodes the full multi-time statistics of the system dynamics with respect to the given intervention time steps.

{For now,} we will consider factorized initial conditions
\begin{equation}
    \rho_\mathrm{tot}=\rho_\mathrm{env}\otimes\rho_0.
\end{equation}
In this case the first intervention can be seen as a preparation step for the system initial state, since the {CP} map $\mathcal{A}_0$ acts on a fixed initial system state $\rho_0$. Thus, we can replace the first intervention by the initial density matrix itself $\mathcal{T}_{t_f,t_M...t_1}[\mathcal{A}_M,...,\mathcal{A}_{1};\rho_0]$. One can then obtain the dynamical map $\mathcal{E}_t$ at timestep $t$ as the process tensor without interventions ($M=1$)
\begin{equation}
    \mathcal{E}_t\rho_0=\mathcal{T}_{t}[\rho_0].
\end{equation}
Such a quantum channel can be identified with a corresponding 
{Choi state via the Choi-Jamiołkowski isomorphism \cite{choi1975Completely,jamiolkowski1972Linear}. The channel $\mathcal{E}_t$ is mapped to a mixed state of a doubled Hilbert-space via
\begin{align}
\label{eq:choi-state}
    \choi[\cpt_{t}] =  (\cpt_{t}\otimes \id_\Anc)[\ketbra{\Psi^+}{\Psi^+}_{\Sys+\Anc}],
\end{align}
where $\ket{\Psi^+_{\Sys+\Anc}}\propto\sum_i\ket{i}_\Sys\otimes\ket{i}_\Anc$ is a maximally entangled state of system and ancilla.} A similar construction exists for process tensors \cite{Pollock2018Jan, taranto2025Higher}, where the Choi state for an $M$-intervention process tensor becomes a many-body mixed state in a Hilbert space of $2M$ copies of the system. 

Both, process tensors and the dynamical map can be used to define notions of memory in the dynamics, including characterizations of quantum or classical memory, as outlined later. For uncorrelated initial conditions, the dynamical map can be inferred directly from multi-intervention process tensors corresponding to the same dynamics, whereas the definition of a dynamical map is ambiguous in the case of initial correlations~\cite{ColNeuBre2022}. {However, it should be stressed that experimentally measuring process tensors via tomography \cite{white2022Non} is much more challenging compared to measuring dynamical maps.}

\section{Process tensors from Matrix Product Operator Influence Matrices}\label{sec:influence_matrix}
Computing process tensors directly from the definition \eqref{eq:Q_def} is not possible in practice if the bath Hilbert space is large. Moreover, standard master equation approaches only determine the dynamical map and simple multi-time correlation functions via the quantum regression theorem. In contrast, numerically exact non-Markovian open system methods can be used to access more general multi-intervention process tensors \cite{Ortega2024Sep}. One class of such exact open system methods aims at computing the so-called {environment influence matrix} \cite{Jorgensen2019Dec, fux2021Efficient,Cygorek2021Jan, Lerose2021May,Ortega2024Sep,Chen2024Jan,Thoenniss2023May, Ng2023Mar,sonner2025Semi}, {an object that encodes the full effect of the environment onto the local system dynamics. By evoking tools from tensor network theory, these algorithms make it possible to automatically generate an effective {compressed auxiliary environment} with a manageable state space that mimics the full environmental influence in a numerically controlled way.} In particular, the uniTEMPO method (uniform time-evolving matrix product operators) presented in Refs.~\cite{Link2024May,kahlert2024Simulating} for Gaussian bosonic baths provides a representation of the process tensor in the form
\begin{equation}\label{eq:PT_MPO}
\begin{split}
        &\mathcal{T}_{t_f,t_M,...,t_1}[\mathcal{A}_M,...,\mathcal{A}_{1};\rho_0]=\\&v_l^T (q)^{t_f-t_M} \mathcal{A}_{M}(q)^{t_{M}-t_{M-1}} \mathcal{A}_{M-1}\cdots (q)^{t_1}(v_r\otimes \rho_0),
\end{split}
\end{equation}
{where $q$ is a linear map on a combined space of system and auxiliary degrees of freedom, as depicted in Fig.~\ref{fig:im}. The boundary vectors $v_{l/r}$ replace $\rho_\mathrm{env}$ and $\tr_\mathrm{env}$ in Eq.~\eqref{eq:pt_def1}. We provide additional information on this formalism in App.~\ref{app:pt}. The objects $q$ and $v_{l/r}$ define a temporal matrix product operator (MPO), which yields a compressed representation of the influence matrix, a time-discrete variant of the Feynman-Vernon influence functional \cite{Feynman1965}. When seen as a MPO, the dimension of the auxiliary space defines the so-called MPO bond-dimension. Since this representation is uniform, i.e.~the map $q$ is the same for each time step, the method provides an embedding of the local dynamics within an {effective dynamical semi-group} \cite{sonner2025Semi}.} 
In general, the required bond dimension depends on the desired accuracy and the specific dynamics. For many systems, including impurity models with continuous baths, the temporal correlations encoded in the influence matrix are limited and the bond dimension required for achieving convergence is low, even at long evolution times \cite{sonner2021Influence,Vilkoviskiy2024May,Nguyen2024Sep,thoenniss2024Efficient}. This property allows us to generate numerically converged process tensors for spin-boson models both in transient and stationary dynamical regimes, i.e.~for arbitrary intervention times $t_m$.

In the context of characterizing the computational complexity of simulating local quantum dynamics, the temporal correlations encoded in the influence matrix are sometimes referred to as ``temporal entanglement" \cite{hastings2015Connecting,lerose2021Influence,sonner2021Influence,foligno2023Temporal,Vilkoviskiy2024May,cerezo2025Spatio}. However, it is important to note that the influence matrix fundamentally represents a mixed state via the Choi representation of a corresponding process tensor. Therefore, a measure for ``temporal entanglement" characterized by the singular values of an orthogonal bond in an MPO influence matrix (for instance the operator space entanglement entropy \cite{wellnitz2022Rise}) describes both quantum and classical temporal correlations. Thus, these measures cannot be used to distinguish quantum and classical memory. For instance, influence matrices arising from purely classical environments, such as random unitary environments or high temperature environments, may have large temporal entanglement \cite{kahlert2024Simulating} even though, by any reasonable definition, they do not represent quantum memory. To characterize quantum memory, one must instead detect proper quantum correlations in the process tensor \cite{TarQuiMurMil2024}. 

\begin{figure}
    \centering
    \begin{tikzpicture}
        \initialbox{0}{0}{$v_r$}{$q$}
        \wirelinedots{2.7}{0}
        \centerbox{2}{0}{$q$}
        \finalbox{4}{0}{$v_l$}{$q$}
    
        \draw[line width=1pt, blur shadow={shadow blur steps=5, shadow xshift=0.3mm, shadow yshift=-0.3mm}, ->] (-1, -1.75) -- node[below] {time} (6, -1.75);
        \draw[decorate, decoration={brace, amplitude=7pt, mirror}, line width=1.05pt] (-0.65, -1.9) -- node[below=6pt] {$\delta t$} (1.35, -1.9);
    \end{tikzpicture}
    
    \begin{tikzpicture}
        \node at (-1,-0.55) {$\mathcal{T}_{t+t_1, t_1} = $};
        \initialbox{1}{0}{$v_r$}{$q$}
        \finalbox{3}{0}{$v_l$}{$q$}
        \node at (0.85, -0.5)  {$\left( \vphantom{\rule{0pt}{1.1cm}} \right.$};
        \node at (1.85, -0.5)  {$\left. \vphantom{\rule{0pt}{1.1cm}} \right)$};
    
        \node at (2.85, -0.5)  {$\left( \vphantom{\rule{0pt}{1.1cm}} \right.$};
        \node at (3.85, -0.5)  {$\left. \vphantom{\rule{0pt}{1.1cm}} \right)$};
        \node at (2, 0.4) {$t_1$};
        \node at (4, 0.4) {$t$};

	\filldraw[line width=1.2pt, rounded corners, fill=white,draw=gray, dashed, blur shadow={shadow blur steps=5, shadow xshift=0.3mm, shadow yshift=-0.3mm}] (2, 0-1.4) rectangle ++(0.7,0.7) node[midway]{$\mathcal{A}_1$};
	\filldraw[line width=1.2pt, rounded corners, fill=white,  draw=gray, dashed, blur shadow={shadow blur steps=5, shadow xshift=0.3mm, shadow yshift=-0.3mm}] (0.2, -1.1) circle (0.3cm) node {$\rho_0$};
    \end{tikzpicture}

    \caption{Upper: Influence matrix (time step $\delta t$) expressed as a temporal uniform matrix product operator. Open legs correspond to free indices that can be contracted with local quantum operations. Lower: Process tensor for a single intervention $\mathcal{A}_1$ after $t_1$ time steps, written as a matrix product operator according to Eq.~\eqref{eq:PT_MPO}.}
    \label{fig:im}
\end{figure}

\section{Quantum memory criteria}\label{sec:qm_crit}

\subsection{Process tensor based criterion}\label{sec:pt_crit}

In this section we introduce the first of two quantum memory criteria that we will use in this work. This criterion is based on the process tensor, specifically single-intervention process tensors $\mathcal{T}_{t_2,t_1}$ ($M=1$). In the case of multiple interventions, a more refined characterization of quantum memory would be possible \cite{MilSpeXuPolModGue2021}, which we do not consider here.
We will make use of the convenient Choi representation of the process tensor \cite{Pollock2018Jan} in the following. 
The Choi state lives in a Hilbert space of several copies of the system. In particular, assuming a system dimension $d$, the Hilbert space for the Choi state of a single-intervention process tensor has dimensions $d^4$, corresponding to initial state, intervention input, intervention output, and final state. The correlation properties of the Choi state with respect to a bipartition into $d^2\times d^2$ can be related to different types of memory in the dynamics \cite{Giarmatzi2021Apr, TarQuiMurMil2024}.
\begin{figure}
    \begin{tikzpicture}
    \node at (-1.4,-1.0) {$\mathcal{T}_{t_2,t_1}= $};
    \node at (-0.32, -1.1) {\Large{$\sum\limits_i$}};
    \qboxs{1}{0}{$\mathcal{I}_i$}{0.75}
    \qboxs{3}{0}{$\Phi_i$}{0.75}
    \qwire{1.75}{-1.1}{0.3}
    \qwire{2.7}{-1.1}{0.3}
    \qwire{0.7}{-1.1}{0.3}
    \qwire{3.75}{-1.1}{0.3}
    \filldraw[line width=1.2pt, rounded corners, fill=white,draw=gray, dashed, blur shadow={shadow blur steps=5, shadow xshift=0.3mm, shadow yshift=-0.3mm}] (2, 0-1.4) rectangle ++(0.7,0.7) node[midway]{$\mathcal{A}$};
    \filldraw[line width=1.2pt, rounded corners, fill=white,  draw=gray, dashed, blur shadow={shadow blur steps=5, shadow xshift=0.3mm, shadow yshift=-0.3mm}] (0.4, -1.1) circle (0.3cm) node {$\rho_0$};
    \end{tikzpicture}\vspace{0.5cm}

\begin{tikzpicture}
\node at (-1,-1) {$\choi\left[\mathcal{T}_{t_2, t_1}\right] = $};
    \node at (0.3, -1.1) {\Large{$\sum\limits_i$}};
    \qboxs{1}{0}{$T_1^i$}{0.75}
    \qboxs{3}{0}{$T_2^i$}{0.75}
    \qwire{1.75}{-1.1}{0.3}
    \qwire{2.7}{-1.1}{0.3}
    \qwire{0.7}{-1.1}{0.3}
    \qwire{3.75}{-1.1}{0.3}
    \node at (2.35, -1.0) {\Large{$\otimes$}};
\end{tikzpicture}
    \caption{Upper: A single-intervention process tensor has classical memory if and only if it can be written as a composition of conditional instruments, as in Eq.~\eqref{eq:classical_memory_pt}. Lower: For all process tensors with classical memory, the Choi state is a separable state according to Eq.~\eqref{eq:classical_memory_pt_sep}.}
    \label{fig:single-intervention-pt}
\end{figure}
A \emph{Markovian} and thus memoryless process is indicated by a process tensor that factorizes, i.e.~the Choi state $\choi$ of a Markovian process can be written as a product state
\begin{equation}\label{eq:pt_crit}
    \choi[\mathcal{T}_{t_2,t_1}]=T_1\otimes T_2
\end{equation}
where the states $T_1, T_2$ are Choi representations corresponding to independent channels on the system. Note that this is a narrow definition of Markovianity. The corresponding dynamical maps satisfy all common notions of Markovianity frequently used in the literature \cite{BreLaiPii2009, BreLaiPiiVac2016, LiHalWis2018, RivHuePle2010, RivHuePle2014}. 

If the Choi state is not a product state we call the dynamics non-Markovian. Such processes with memory can be divided further into two subclasses, processes with \emph{classical memory} and processes with \emph{quantum memory}.

{A process tensor has classical memory if it can be written as a composition of conditioned instruments \cite{Giarmatzi2021Apr, TarQuiMurMil2024}. In the case of a single-intervention $\mathcal{A}_1$, this means that for a classical memory process
\begin{align}
    \label{eq:classical_memory_pt_def0}
    \mathcal{T}_{t_2,t_1}[\mathcal{A}_1;\rho_0]=\sum_{i,j} \mathcal{I}_j^{(i)}  \mathcal{A}_1 \mathcal{I}_i^{} \rho_0
\end{align}
with $\mathcal{I}_i$ and $\mathcal{I}_j^{(i)}$ being quantum instruments on the system, where the latter one is conditioned on the choice of the instrument $\mathcal{I}_i$. Note that, in Eq.~\eqref{eq:classical_memory_pt_def0}, the sum over the second set of instruments can be performed to obtain the conditioned CPT maps $\Phi_i=\sum_j\mathcal{I}_j^{(i)}$, such that the expression simplifies to 
\begin{align}
    \label{eq:classical_memory_pt}
    \mathcal{T}_{t_2,t_1}[\mathcal{A}_1;\rho_0]=\sum_{i} \Phi_i \mathcal{A}_1 \mathcal{I}_i \rho_0,
\end{align}
displayed in Fig.~\ref{fig:single-intervention-pt}. For the goal of confirming quantum memory, it is convenient to consider a broader class of process tensors given by the process tensors with separable Choi state (class SEP in Ref.~\cite{TarQuiMurMil2024})
\begin{equation}
    \label{eq:classical_memory_pt_sep}
    \choi[\mathcal{T}_{t_2,t_1}]=\sum_i T_{1,i}\otimes T_{2,i}.
\end{equation}
Processes with classical memory form a strict subset of this class, i.e.~all classical memory process tensors have a separable Choi state.
This allows us to use methods from entanglement detection to falsify the hypothesis of purely classical memory.} In particular, processes for which the Choi state is entangled must have \emph{quantum memory} because they cannot be represented in form of Eq.~\eqref{eq:classical_memory_pt}. 
{Determining whether an arbitrary state is entangled is an NP-hard problem \cite{Hua2014}. We will thus use a computationally accessible lower bound $\conc_<$ of the so-called I-concurrence as a sufficient criterion to witness entanglement \cite{CheAlbFei2005, Uhl2000, RunBuzCavHilMil2001, Wootters1998}, see App.~\ref{sec:ptcriterion_details} for details.} 
Using the criterion
\begin{equation}\label{eq:conc_main_text}
    \mathcal{C}_<[\chi]>0
\end{equation} 
we can investigate entanglement properties of higher-dimensional states, such as the process tensor Choi states $\chi$, and thus detect violations of classical memory conditions via falsification of the representation~\eqref{eq:classical_memory_pt_sep}.

\subsection{Map-based criterion}

While the process tensor provides the most general framework for characterizing local quantum dynamics, {the existence of quantum memory in the environment} can also be revealed using only the dynamical map $\cpt_t$ evaluated at two evolution times. This is achieved through a definition for dynamics with quantum memory introduced in Ref.~\cite{Backer2024Feb}, which we briefly recapitulate in the following. 
We call the two CPT maps $\cpt_{t_1}$ and $\cpt_{t_2}$ a \emph{dynamics} $\Dyn$ (we require $t_1<t_2$). 
A dynamics 
is \emph{realizable with classical memory} if there exists a set of Kraus operators $\{ K_i\}$ and a set of CPT maps $\Phi_i$ such that 
\begin{align}\label{eq:map_def}
        \cpt_{t_1}\rho = \sum_i K_i \rho K_i^\dagger, && \cpt_{t_2}\rho = \sum_i \Phi_i (K_i \rho K_i^\dagger).
\end{align}
Otherwise the dynamics is said to require quantum memory.
Dynamics with classical memory according to this definition can be realized  by the following classical feedback protocol: First one performs a measurement on the system such that an average over all outcomes realizes the map $\cpt_{t_1}$. Now, conditioned on the outcome $i$ of this measurement, one applies a CPT map $\Phi_i$. Upon discarding the measurement information this procedure yields the overall map $\cpt_{t_2}$ by assumption. Crucially, storing the measurement outcome $i$ requires classical memory only, and the subsequent evolution given by the conditioned CPT map $\Phi_i$ can be realized with a new, uncorrelated environment. 
Related constructions can be found in Refs.~\cite{LiHalWis2018, TarQuiMurMil2024}.

We can establish a connection of the definition of classical memory based on the quantum dynamical map to the definition of classical memory dynamics based on the process tensor. 
First, it is important to note that the pair of maps contains less information on the local dynamics compared to the single-intervention process tensor. 
In fact, the two channels $\cpt_{t_1}$ and $\cpt_{t_2}$ can be obtained directly from the single-intervention process tensor $\mathcal{T}_{t_2,t_1}$, as illustrated in Fig.~\ref{fig:pt_to_maps}. Yet, many different process tensors may exist that realize the same pair of channels. 
The classical feedback protocol realizing Eq.~\eqref{eq:map_def} defines {and hence guarantees the existence of at least one} process with classical memory according to the process tensor criterion Eq.~\eqref{eq:classical_memory_pt} {via $\mathcal{I}_i\rho=K_i \rho K_i^\dagger$.} If a process tensor obeys the classical memory condition \eqref{eq:classical_memory_pt} it thus also follows directly that the corresponding dynamics can be realized with classical memory as in Eq.~\eqref{eq:map_def}.
On the other hand, given a process tensor with quantum memory according to the process tensor definition, a corresponding dynamics ($\cpt_{t_1}$,~$\cpt_{t_2}$) may still be realizable with classical memory. {This hierarchy directly reflects the different information content of dynamical maps and process tensors. Process tensors encode more information, making them harder to measure but also providing stronger signatures of quantum memory present in the environment.}

Criteria to detect quantum memory via Eq.~\eqref{eq:map_def} have been provided in Refs.~\cite{Backer2024Feb, backer2025Entropic}. Here, we introduce and propose an alternative criterion based on the I-concurrence. In our numerical calculations, we found that this is a more powerful criterion for the {two-spin-boson model}, compared to the criteria used in previous works \cite{backer2025Entropic}.
We first compute the Choi states $\chi_{t_i} \equiv \choi[\cpt_{t_i}]$ of the maps $\cpt_{t_1}$ and $\cpt_{t_2}$ {by their local action on a bipartite state of system and ancilla via Eq.~\eqref{eq:choi-state}}.
We then define
\begin{align}
    \label{eq:definition_delta}
    \Delta(t_1, t_2) =\max \left\{0,\conc_{<}(\chi_{t_2})-\sqrt{2 \left(1 - \tr \left(\tr_{\Anc}(\chi_{t_1})^2\right)\right)}\right\}
\end{align}
with $\conc_<$ from Eq.~\eqref{eq:lower_bound_iconcurrence} and using Ref.~\cite{LiFeiAlbLiu2009}
and formulate the following criterion:
Given two maps $\cpt_{t_1}$ and $\cpt_{t_2}$, if for the corresponding Choi states $\choi_{t_1}$ and $\choi_{t_2}$ we observe that
\begin{align}\label{eq:map_crit_witness}
    \Delta(t_1, t_2) >0,
\end{align}
with $\Delta$ as defined in Eq.~\eqref{eq:definition_delta}, the dynamics $\Dyn = \dynamics{\cpt_{t_1}, \cpt_{t_2}}$ is not realizable with classical memory according to Eq.~\eqref{eq:map_def}. 
A proof of this statement and further details on the broader applicability of this criterion are provided in App.~\ref{sec:mapcriterion_details}. {Note that in the case of a two-level system the criterion from Ref.~\cite{Backer2024Feb} turns out to be slightly superior, but it cannot be applied easily beyond the qubit case.}

\begin{figure}
    \centering
    \begin{tikzpicture}[scale=1]
        \node at (-0.6,-0.55) {$\mathcal{E}_{t_2} = $};
        \initialbox{1}{0}{$v_r$}{$q$}
        \finalboxcustom{3}{0}{$v_l$}{$q$}{1}
        \node at (0.85, -0.5)  {$\left( \vphantom{\rule{0pt}{1.1cm}} \right.$};
        \node at (1.85, -0.5)  {$\left. \vphantom{\rule{0pt}{1.1cm}} \right)$};
    
        \node at (2.85, -0.5)  {$\left( \vphantom{\rule{0pt}{1.1cm}} \right.$};
        \node at (3.85, -0.5)  {$\left. \vphantom{\rule{0pt}{1.1cm}} \right)$};
        \node at (2.1, 0.4) {$t_1$};
        \node at (4.1, 0.4) {$t$};

	\begin{scope}[shift={(0,-3)}]
        \node at (-0.6,-0.55) {$\mathcal{E}_{t_1} = $};
        \initialbox{1}{0}{$v_l$}{$q$}
        \qwirev{2}{-1.08}{-0.75}
        \finalbox{3}{0}{$v_r$}{$q$}
        \node at (0.85, -0.5)  {$\left( \vphantom{\rule{0pt}{1.1cm}} \right.$};
        \node at (1.85, -0.5)  {$\left. \vphantom{\rule{0pt}{1.1cm}} \right)$};
    
        \node at (2.85, -0.5)  {$\left( \vphantom{\rule{0pt}{1.1cm}} \right.$};
        \node at (3.85, -0.5)  {$\left. \vphantom{\rule{0pt}{1.1cm}} \right)$};
        \node at (2.1, 0.4) {$t_1$};
        \node at (4.1, 0.4) {$t$};
        \node at (4.5, -1.1) {tr};
        \node at (2.5, -1.1) {$\sigma_0$};
    \end{scope}
    \end{tikzpicture}
    \caption{Extraction of the two dynamical maps $\mathcal{E}_{t_1}$ and $\mathcal{E}_{t_2}$ from the single-intervention process tensor $\mathcal{T}_{t_2,t_1}$ (see Fig.~\ref{fig:im}, $t_2=t_1+t$). For the map at $t_2$ one simply connects the outgoing and ingoing legs of the intervention at $t_1$. To recover the intermediate map $\mathcal{E}_{t_1}$ one leaves the outgoing leg open and contracts the ingoing leg with an arbitrary state $\sigma_0$, traced out at the end.} 
    \label{fig:pt_to_maps}
\end{figure}

\section{Spin-boson model}\label{sec:spin_boson}

The spin-boson model is a standard example for studying dissipative quantum dynamics outside the Markovian regime, and it is used commonly as a benchmark for exact and approximate numerical open system methods \cite{strathearn2018Efficient,hartmann2020Accuracy,xu2022Taming,iles2024Capturing,le2024Managing,dowling2024Capturing}. The model consists of a single spin (``impurity") coupled to a reservoir of noninteracting bosonic modes. The Hamiltonian can be written as
\begin{equation}
    H=\Omega S_x+S_z \sum_\lambda g_\lambda (b_\lambda+b_\lambda^\dagger )+\sum_\lambda \omega_\lambda b_\lambda^\dagger b_\lambda,
\end{equation}
where $S_{x/y}$ denote components of the systems spin operator $\Vec{S}$ and $b_\lambda$ are bosonic annihilation operators for individual bath modes. The bath structure can be characterized by a spectral density
\begin{equation}
    J(\omega)={\pi}\sum_\lambda |g_\lambda|^2\delta(\omega-\omega_\lambda).
\end{equation}
Usually one assumes a continuum of modes leading to a smooth spectral density and a finite bath memory time. We restrict our simulations to single spin $\Vec{S}=\frac{1}{2}\Vec{\sigma}$ and two-spin $\Vec{S}=\frac{1}{2}(\Vec{\sigma}_1+\Vec{\sigma}_2)$ models.

For now, we consider quench dynamics, starting with an uncoupled bath in the bare vacuum state (zero temperature), and switching on the coupling to the system at time $t_i=0$. Note that both the dynamical map (Fig.~\ref{fig:2}) and the single-intervention process tensor (Fig.~\ref{fig:1}) do not require the specification of an initial state for the system. We assume moderate coupling strengths so that our numerics converges reliably. An example for a convergence analysis is provided in App.~\ref{app:conv}.

\subsection{Lorentzian spectral density}
As a first class of environments we consider a Lorentzian spectral density
\begin{equation}
    J(\omega)=\frac{\eta\gamma}{\gamma^2+(\omega_0-\omega)^2}.
\end{equation}
In the limit $\gamma\rightarrow 0$ this reduces to a single mode in the bath with Hamiltonian \cite{hartmann2020Accuracy}
\begin{equation}
    H=\Omega S_x+\sqrt{\eta}S_z (b+b^\dagger)+\omega_0b^\dagger b,
\end{equation}
i.e.~fully coherent dynamics akin to previously studied examples, see for instance Ref.~\cite{Giarmatzi2021Apr}. The spectral broadening due to a finite value of $\gamma$ introduces a finite memory time ($\propto 1/\gamma$) to the environment.

\begin{figure}
    \centering
    \includegraphics[width=0.85\linewidth]{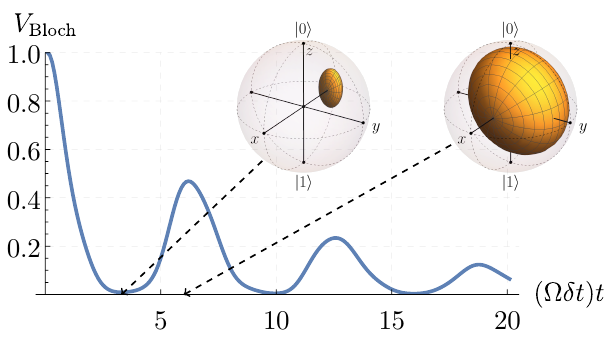}
    \caption{Deformation of the Bloch sphere under the dynamics of the one-spin-boson model with Lorentzian bath with parameters $\gamma=0.075\Omega$, $\omega_0=1.5\Omega$, $g=\Omega$. The specific maps depicted in the inserts correspond to the two time points for which quantum memory is maximal according to the map criterion in the upper left panel in Fig.~\ref{fig:1}. The coherent dynamics leads to strong revivals in the Bloch sphere volume which violate all common notions of Markovianity.}
    \label{fig:bloch_lorentz}
\end{figure}

\begin{figure}
    \centering
    \includegraphics[width=0.48\linewidth]{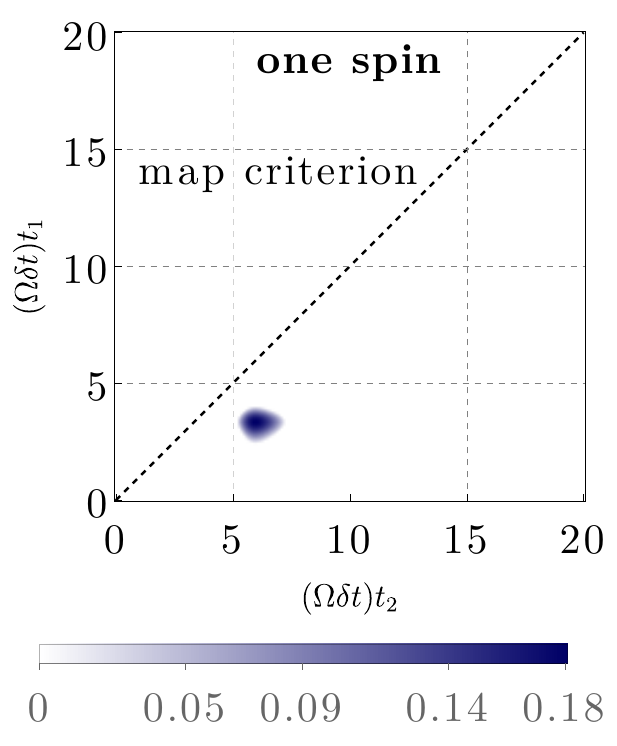}
    \includegraphics[width=0.48\linewidth]{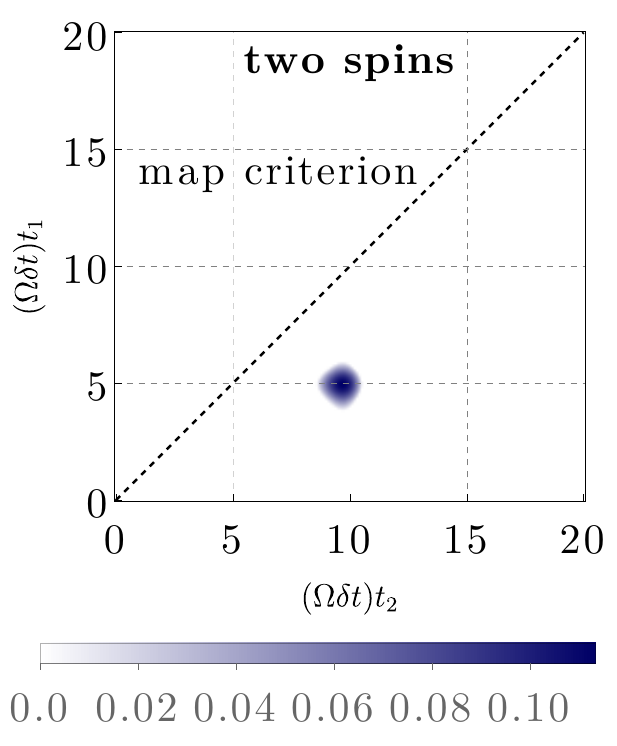}\\
    \includegraphics[width=0.48\linewidth]{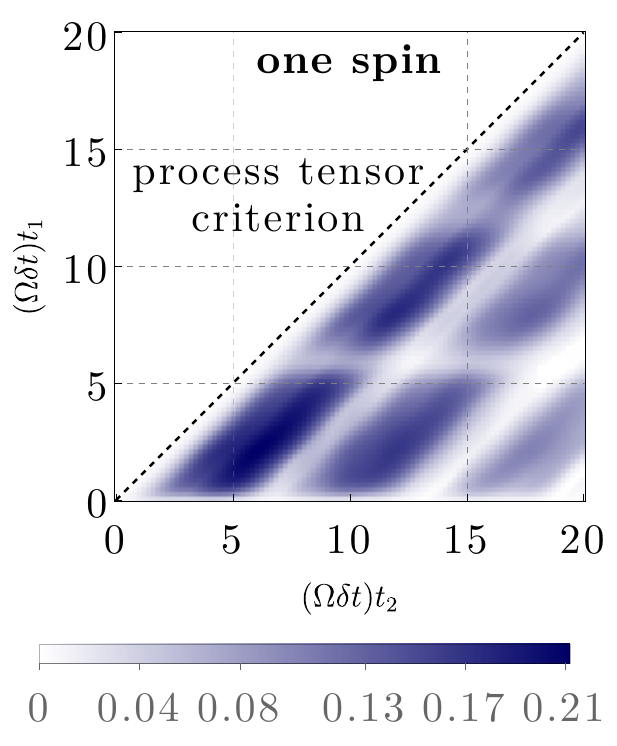}
    \includegraphics[width=0.48\linewidth]{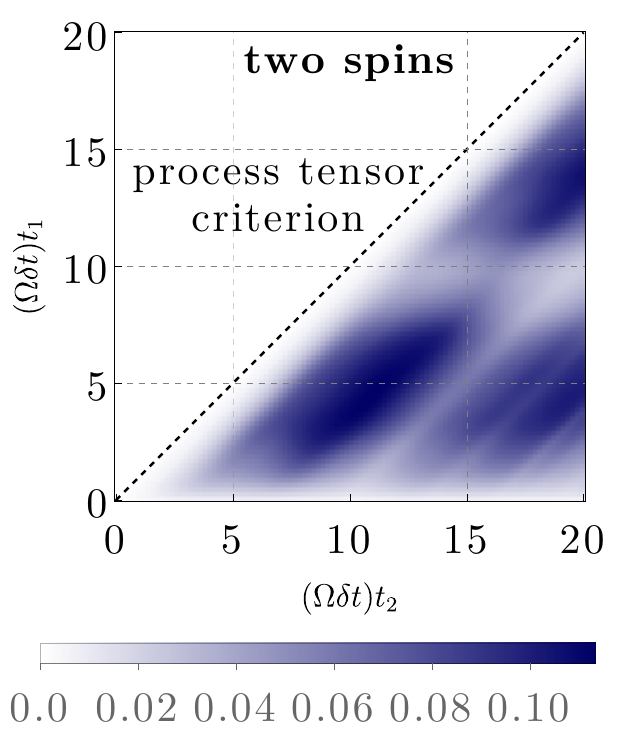}
    \caption{Quantum memory criteria for the spin-boson model with Lorentzian bath (left: single spin, $\gamma=0.075\Omega$, $\omega_0=1.5\Omega$, $g=\Omega$; right: two spins, $\gamma=0.05\Omega$, $\omega_0=1.5\Omega$, $g=\Omega$). The figures show the quantities $\Delta(t_1,t_2)$ from Eq.~\eqref{eq:map_crit_witness} (top) and $C_<[\chi]$ from Eq~\eqref{eq:conc_main_text} (bottom) in color-code (color label below each figure). Positive values (blue color) imply quantum memory for the two times $(t_1,t_2)$, respectively. For the chosen fine-tuned parameters, both criteria find quantum memory.}
    
    \label{fig:1}
\end{figure}

For the numerical simulations that we display in the following we used fine-tuned parameters chosen such that the map criterion can predict quantum memory. This required small $\gamma$ (long memory times) and $\omega_0$ close to resonance. Otherwise the criterion was not able to indicate quantum memory for any combination of times $t_1$ and $t_2$. {To first get a feeling for the dynamics we consider the single spin case, which can be conveniently displayed via the deformation of the Bloch sphere under the action of the dynamical map. We show in Fig.~\ref{fig:bloch_lorentz} the Bloch sphere volume as a function of the evolution time. One can see that the map indicates strong non-Markovianity, with the Bloch-sphere volume shrinking and increasing periodically, a signature of P-indivisibility \cite{BreLaiPii2009, LorPlaPat2013}. Due to the resonant environment, excitations are exchanged coherently with the bath leading to periodic revivals in the Bloch-sphere volume.} In Fig.~\ref{fig:1} we computed both the process tensor criterion and the map criterion for quantum memory for a quench with the Lorentzian bath. 
In the plots, we display $\Delta(t_1, t_2)$ and the entanglement criterion for the process tensor (lower bound $\conc_<$ of the I-concurrence) with intervention times $t_1<t_2$. Baths with finite memory time lead to a relaxation of the local state such that the map criterion can only be fulfilled at short times, see section \ref{sec:stat_qm}. {The map criterion only predicts quantum memory in a small window of times during the first revival, close to the points where the increase of the Bloch sphere volume displayed in Fig.~\ref{fig:bloch_lorentz} is maximal.}  On the other hand, the process tensor criterion reliably predicts quantum memory in broad dynamical regimes, and also for short memory times. {The pattern seen in $\conc_<$ aligns with the periodicity observed in the dynamics in Fig.~\ref{fig:bloch_lorentz}.} It is important to keep in mind that the quantities displayed in Fig.~\ref{fig:1} are lower bounds that do not quantify the amount of quantum memory in an operational way. 
Inequalities~\eqref{eq:map_crit_witness} (top) and~\eqref{eq:conc_main_text} (bottom) are sufficient, but not necessary criteria for quantum memory. Thus, while a positive value (blue color) verifies the necessity of quantum memory, the value zero (white) does not imply that the dynamics can be realized with classical memory.

\subsection{Ohmic spectral density}

\begin{figure}
    \centering
    \includegraphics[width=0.85\linewidth]{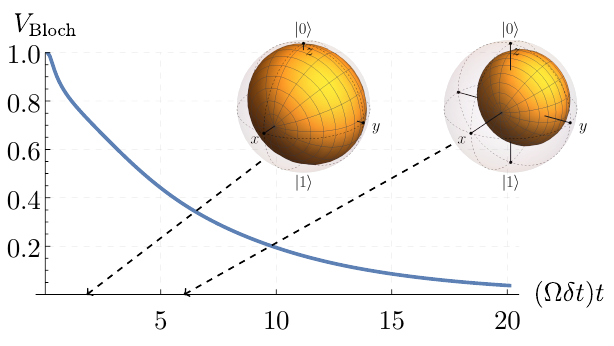}

    \caption{Deformation of the Bloch sphere under the dynamics of the one-spin-boson model with Ohmic bath $\alpha=0.2$, $\omega_c=5\Omega$ at zero temperature. The evolution reflects the fact that the entanglement of the Choi state \cite{RivHuePle2010} as well as the volume of the Bloch ball \cite{LorPlaPat2013} is always decreasing. In contrast to the more coherent dynamics shown in Fig.~\ref{fig:bloch_lorentz}, the dynamical map is P-divisible, but still CP-indivisible.} 
    \label{fig:bloch_ohm}
\end{figure}

\begin{figure}
    \centering
    \includegraphics[width=0.48\linewidth]{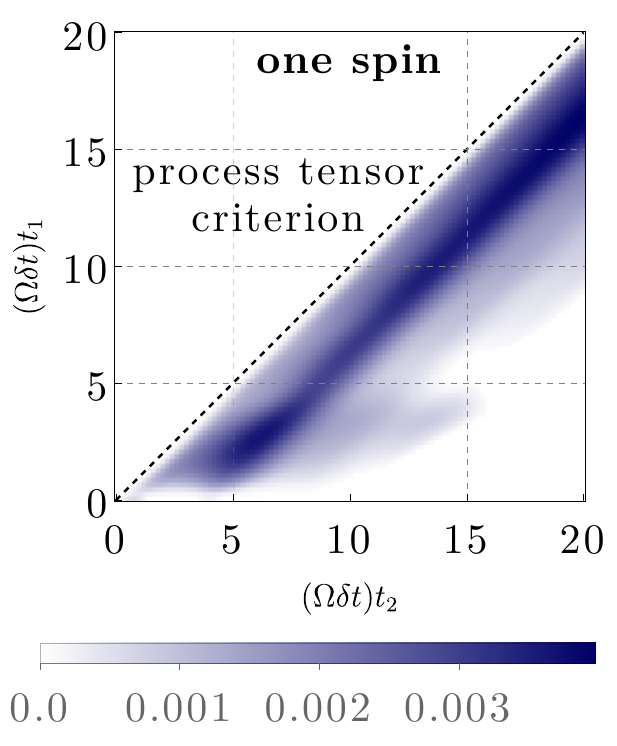}
    \includegraphics[width=0.48\linewidth]{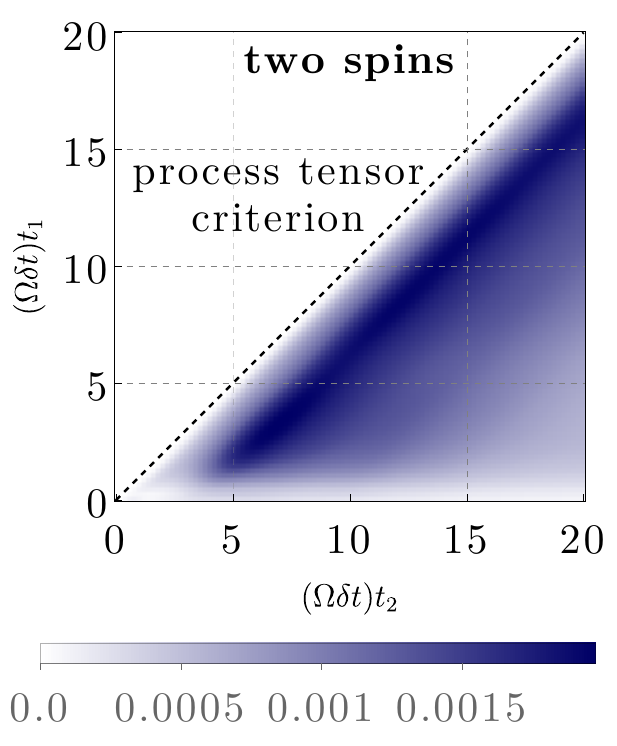}
    \caption{Quantum memory criterion for spin-boson dynamics, as in Fig.~\ref{fig:1}, but with an Ohmic bath $\alpha=0.2$, $\omega_c=5\Omega$ at zero temperature (left: single spin, right: two spins).  The criterion based on the dynamical map fails to predict quantum memory in this case.}
    \label{fig:2}
\end{figure}

We now consider Ohmic baths with exponential high-frequency cutoff, characterized by the spectral density
\begin{equation}
    J(\omega)=\Theta(\omega)\alpha \omega \eul^{-\omega/\omega_c},
\end{equation}
where $\Theta$ is the step function and $\omega_c$ is a high frequency cutoff. The strong low-frequency components of the bath spectrum induce algebraic correlations in this model that are absent for the simpler Lorentzian bath. In contrast to the Lorentzian spectral density, which features negative frequency components, Ohmic baths are well-defined at all temperatures, and lead to an equilibration in the local dynamics. {We again display the Bloch-sphere deformation in the single-qubit case in Fig.~\ref{fig:bloch_ohm}. In contrast to the Lorentzian bath dynamics from Fig.~\ref{fig:bloch_lorentz}, the Bloch sphere volume decreases monotonically and the map is P-divisible. However, the dynamics is still CP-indivisible \cite{RivHuePle2010, HalCreLiAnd2014}, i.e.~it cannot be explained by a strictly Markovian process. However, we could not find a parameter regime that lead to a positive prediction of quantum memory via the map criterion, even at zero temperature.} We attribute this to a generally lower level of quantum memory in the process when compared with the more coherent dynamics shown in Fig.~\ref{fig:1}. The broad Ohmic spectrum results in a relatively smooth and incoherent evolution to equilibrium, such that it is difficult to indicate quantum memory given only the time-local information encoded in the dynamical map.
On the other hand, the criterion based on the process tensor does predict quantum memory in the dynamics, as displayed in Fig.~\ref{fig:2}. This behavior persists beyond the zero-temperature case, as discussed in the next section. The dynamics with an Ohmic bath quickly reaches a stationary regime. Therefore, quantum memory vanishes exactly if the time difference $t_2-t_1$ is large{, explaining the decay of $\conc_<$ along the anti-diagonal in  Fig.~\ref{fig:2}}. The behavior of the process tensor criterion is similar for the single- and two-spin models. For the considered parameters, equilibration in the two-spin model is slower, such that the criterion still predicts quantum memory at larger time differences.

\section{quantum memory in the stationary dynamics}\label{sec:stat_qm}

{For characterizing quantum non-Markovian effects, most investigations have been focused on revivals in the transient dynamics encoded in the dynamical map \cite{BreLaiPii2009, RivHuePle2010, BreLaiPiiVac2016, BuscemiPRXQuantum2025}. However, in many realistic scenarios the system is initially already in equilibrium with the environment. We can think of this stationary regime as emerging from an equilibration of an uncorrelated state, i.e.~large times $t_1,t_2\rightarrow\infty$. In this view the channels $\cpt_{t_1}$ and $\cpt_{t_2}$ become identical maps of any input state to a unique steady state density matrix and the dynamics $\mathcal{D}$ becomes memoryless, so that there is trivially no quantum memory. }

{In order to systematically study quantum memory effects in this case, process tensors provide an appropriate framework, as they encode more general multi-time statistics. They can be used to describe situations where the system is forced out of equilibrium by a local intervention, such as a local measurement, triggering a subsequent relaxation dynamics. Thus, the classical memory condition for the process tensor \eqref{eq:classical_memory_pt} can still be violated at large evolution times. In the single-intervention process tensor (Fig.~\ref{fig:im}), the stationary regime can be obtained by considering a finite $t=t_2-t_1$ and formally setting $t_1\rightarrow \infty$. With the MPO form of the process tensor obtained through uniTEMPO (Eq.~\eqref{eq:PT_MPO}), the long-time limit can be performed explicitly  
\begin{equation}\label{eq:PT_MPO_asym}
\begin{split}
        \mathcal{T}_{t+t_1,t_1}[\mathcal{A}_{1};\rho_0]&=v_l^T (q)^{t} \mathcal{A}_{1}(q)^{t_1}(v_r \otimes\rho_0)\\&\xrightarrow{t_1\rightarrow \infty} v_l^T (q)^{t} \mathcal{A}_{1}\Gamma_0 \,\tr\rho_0.
\end{split}
\end{equation}
$\Gamma_0$ is the leading left eigenvector of the effective dynamical propagator $q$ ($q\Gamma_0=\Gamma_0$), which can be efficiently determined using Krylov methods. Thus, due to the effective dynamical semi-group defined via the influence matrix MPO, it is possible to directly realize the proper stationary initial conditions without explicit evolution to long times \cite{Link2024May,sonner2025Semi}. Note that the stationary process tensor now descibes dynamics with initial correlations due to the non-factorized initial state. In the case of the spin-boson model, this state represents the full equilibrium state at the given bath temperature \cite{Nguyen2024Sep}.}

\begin{figure}
    \centering
    \includegraphics[width=0.8\linewidth]{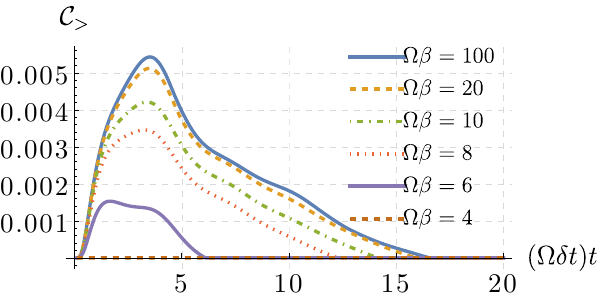}
    \includegraphics[width=0.8\linewidth]{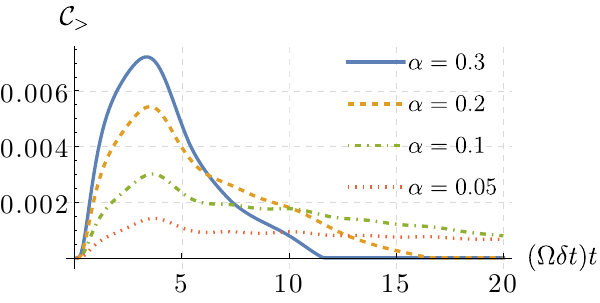}
    \includegraphics[width=0.8\linewidth]{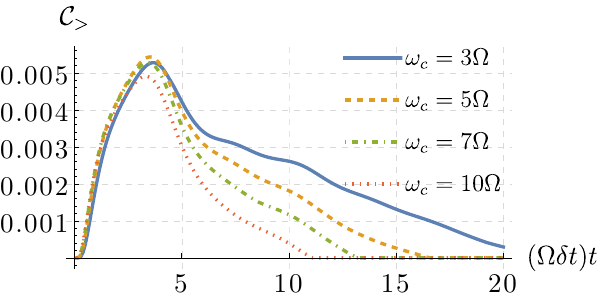}
    \caption{Quantum memory criterion based on the process tensor in the stationary dynamics of the single spin-boson model with different Ohmic baths. $t$ is the time step difference between interventions at asymptotic times. Upper: Varying temperatures $k_\mathrm{B}T=1/\beta$ at $\alpha=0.2$, $\omega_c=5\Omega$. Center: Varying coupling strength at $\beta=\infty$, $\omega_c=5\Omega$. Lower: Varying high-frequency cutoffs $\omega_c$ at $\beta=\infty$, $\alpha=0.2$.}
    \label{fig:3}
\end{figure}

We can now apply the process tensor quantum memory criterion for different intervention time delays $t$. Our results for Ohmic baths are depicted in Fig.~\ref{fig:3}. As expected, quantum memory vanishes at large time differences where the process tensor factorizes due to the loss of memory in the environment. Moreover, the lower bound $\conc_<$ decreases monotonically with increasing temperature $k_\mathrm{B}T=1/\beta$. In the spin-boson model, finite thermal fluctuations can be seen as random unitary noise \cite{Feynman1965,kahlert2024Simulating} and, as such, cannot serve as quantum memory. At high temperature thermal fluctuations dominate the dynamics and the bath memory becomes classical. {With the parameters used in Fig.~\ref{fig:3} our criterion can no longer predict quantum memory for temperatures beyond $\Omega\beta\approx 4$}. The dependence of the quantum memory criterion on the bath coupling strength (Fig.~\ref{fig:3} center panel) is more complex. While the maximum value for the lower bound on the I-concurrence increases with the coupling strength, a stronger coupling leads to faster relaxation and hence to a decrease in the bath memory time.  {Note that, in our simulations, the coupling strength remains below the quantum phase transition of the spin-boson model \cite{vojta2006Impurity}, at which the steady-state becomes degenerate and numerical simulations become much more challenging \cite{strathearn2018Efficient}.} Lastly, increasing the high-frequency cutoff $\omega_c$ does not significantly alter the maximum value of $\conc_<$ predicted by the bound, but merely leads to a slower decay due to the slower bath dynamics.

\section{Conclusions}\label{sec:conclusions}
We studied quantum memory in the dynamics of spin-boson models using different criteria. With the single-intervention process tensor, the existence of quantum memory can generally be confirmed at low temperatures.
In contrast, a different criterion based on the dynamical map alone identifies quantum memory in spin-boson models only when the bath features a sharp resonance, i.e.~when the dynamics is more coherent, and fails entirely in the case of Ohmic baths. This discrepancy arises due to the fact that the dynamical map contains strictly less information compared to process tensors. {In fact, a pair of channels that can be realized by classical-memory dynamics may still be the marginal of a process tensor with quantum memory. If on the contrary the maps do not admit a classical realization then there exists no compatible classical-memory process tensor. Within our current analytical bounds, the time-local information provided by dynamical maps is not sufficient to reveal the quantum memory for spin boson models in common parameter regimes.}

{When considering the stationary regime, a dynamical map cannot be used because the initial state is no longer uncorrelated. In this case the process tensor provides an appropriate framework with which we are able to confirm quantum memory at short times after an intervention.} In experiments, criteria based on dynamical maps are more easily accessible, as fewer measurements are required for tomography of the map compared with the process tensor. It would thus be desirable to find a way to modify the dynamics such that the map-based criterion indicates quantum memory in broader parameter regimes. One idea in this direction could be to use a time-dependent driving on the system, optimized to maximize the quantum memory criterion Eq.~\eqref{eq:map_crit_witness}. In fact, such optimal control problems can be implemented very efficiently using numerical methods based on influence matrices \cite{fux2021Efficient,Ortega2024Sep}. 
We used analytical lower bounds for the I-concurrence in order to construct analytical criteria for temporal quantum correlations. 
With more refined numerical schemes for detection of entanglement, such as semidefinite programming, it could be possible to obtain more sensitive \cite{Giarmatzi2021Apr,yu2025Quantum} or more easily accessible \cite{santos2025Two} tests. We considered exclusively pairs of dynamical maps and single-intervention process tensors. The investigation of multi-intervention process tensors based on multipartite entanglement leads to a refinement of the memory classification in the dynamics \cite{TarQuiMurMil2024}, in accordance with different classes of multipartite entanglement~\cite{MilSpeXuPolModGue2021}. 

Accurate computations of process tensors are made possible through a highly efficient real-time solver for spin-boson models, based on a uniform matrix product operator representation of the influence matrix \cite{Link2024May}. 
In the future, these calculations could also be performed for fermionic impurity models, for which a similar algorithm has recently been introduced \cite{sonner2025Semi}. Moreover, the influence matrix approach can also be applied to interacting environments \cite{banuls2009Matrix,hastings2015Connecting,lerose2021Influence,sonner2021Influence,frias2022Light,foligno2023Temporal}. One interesting question in connection to this would be how interactions in the bath affect the quantum memory. In generic chaotic systems, the required bond dimension of the influence matrix, as quantified by the ``temporal entanglement", increases exponentially with time \cite{foligno2023Temporal}. For an assessment of the fundamental complexity of the local dynamics, it would be interesting to investigate whether these strong temporal correlations are classical or quantum.
\vspace{0.3em}

\section*{Acknowledgements}
The authors thank Konstantin Beyer for insightful discussions and valuable feedback regarding the manuscript. We are grateful to Mei Yu for helpful conversations. CB acknowledges support by the German Academic Scholarship Foundation. The Article Processing Charges were funded by the joint publication funds of the TU Dresden and the SLUB Dresden as well as the Open Access Publication Funding of the DFG. VL acknowledges supported by the DFG via the Research Unit FOR 5688 (Project No. 521530974).

\bibliography{bib.bib}

\appendix

\section{From the influence matrix to the process tensor}\label{app:pt}
The influence matrix formulation of open quantum systems is based on a discretized path integral representation of the system dynamics. We consider specifically the Hamiltonian 
\begin{equation}\label{eq:hamilt_2}
    H=H_\mathrm{sys}+H_\mathrm{env},
\end{equation}
\begin{equation}\label{eq:H_int}
    H_\mathrm{env}=S\sum_\lambda g_\lambda (b_\lambda +b_\lambda^\dagger)+\sum_\lambda \omega_\lambda b_\lambda^\dagger b_\lambda,
\end{equation}
with self-adjoint system coupling operator $S$.
For simplicity, we work in the eigenbasis of this operator $S\ket{n}=S_n\ket{n}$ and introduce a superoperator notation by vectorizing operators as $O^{(n,m)}\equiv\braket{n|O|m}$. Greek indices are used to label all double-index combinations $\mu\equiv (n,m)$. In order to define the influence matrix we consider the unitary channel generated by the environment Hamiltonian \eqref{eq:H_int} for a small time step $\delta t$. This channel is diagonal in the coupling operator eigenbasis, so that we can write
\begin{equation}
    \mathcal{U}_\mathrm{env}=\sum_\mu \tilde{\mathcal{U}}^{(\mu)}_\mathrm{env} \otimes (\ketbra{n}{n}\otimes \ketbra{m}{m}).
\end{equation}
Analogously, we define local superoperators $\tilde{\mathcal{U}}_\mathrm{sys}^{(\mu)}$ via
\begin{equation}
    \tilde{\mathcal{U}}_\mathrm{sys}^{(\mu),\alpha\beta}=\mathcal{U}_\mathrm{sys}^{\alpha\mu}\,\mathcal{U}_\mathrm{sys}^{\mu\beta},
\end{equation}
where $\mathcal{U}_\mathrm{sys}$ is the local channel generated by $H_\mathrm{sys}$ for a half time step $\delta t/2$. Using symmetric Trotter splitting with respect to $H_\mathrm{sys}$ and $H_\mathrm{int}$ ($\mathcal{Q}\approx\mathcal{U}_\mathrm{sys}\mathcal{U}_\mathrm{env}\mathcal{U}_\mathrm{sys}$), the dynamics of the system state for $N$ time steps can be expressed compactly as
\begin{equation}
    \rho_t =\sum_{\mu_1...\mu_N}\mathcal{F}^{\mu_N...\mu_1} \tilde{\mathcal{U}}_\mathrm{sys}^{(\mu_N)}\cdots \tilde{\mathcal{U}}_\mathrm{sys}^{(\mu_1)}\rho_0,
\end{equation}
where the influence matrix $\mathcal{F}$ (time discrete Feynman-Vernon influence functional) is given as
\begin{equation}
    \mathcal{F}^{\mu_N...\mu_1}=\tr_\mathrm{env}\tilde{\mathcal{U}}_\mathrm{env}^{(\mu_N)}\cdots \tilde{\mathcal{U}}_\mathrm{env}^{(\mu_1)} \rho_\mathrm{env}.
\end{equation}
For Gaussian bosonic environments, as in Eq.~\eqref{eq:H_int}, the algorithm presented in Ref.~\cite{Link2024May} can be used to generate a uniform MPO representation of the influence matrix
\begin{equation}
    \mathcal{F}^{\mu_N...\mu_1}={v}_l^T f^{\mu_N}\cdots f^{\mu_1}{v}_r.
\end{equation}
The dimension of the matrices $f^\mu$ (MPO bond dimension) depends on the desired accuracy of the compression. The vectors $v_{l/r}$ realize the desired boundary conditions. We use the scheme from Ref.~\cite{Link2024May} to find these vectors for a quench from an uncoupled bath.  The uniform MPO representation can be used to define an effective semi-group for the dynamics via the auxiliary propagator \cite{Link2024May,sonner2025Semi}
\begin{equation}
    q_{r\alpha,s\beta}=\sum_\mu f^\mu_{rs}\,\tilde{\mathcal{U}}^{(\mu),\alpha\beta}_\mathrm{sys}.
\end{equation}
In the (trotterized) path integral formalism, the inclusion of interventions on the system is straightforward \cite{Jorgensen2019Dec}. Process tensors with arbitrary interventions can thus be obtained numerically from Eq.~\eqref{eq:PT_MPO}.

\section{Details on the process tensor criterion}
\label{sec:ptcriterion_details}

Although the separability problem is NP-hard, there are sufficient criteria for entanglement making it possible to detect quantum memory with low numerical effort.
In this article we choose a lower bound of the I-concurrence of the Choi state as our entanglement criterion. The I-concurrence is a generalization of the well-known single qubit concurrence \cite{Wootters1998} to higher-dimensional states. 
It can be constructed directly from the pure state concurrence in higher dimensions \cite{Uhl2000, RunBuzCavHilMil2001} via a convex roof extension
\begin{align}
    \label{eq:def_i_concurrence}
    \conc[\chi] = \min_{\{p_k, \ket{\psi_k}\}} \sum_{k}^{} p_k \sqrt{1-\tr_1\left((\tr_2\ketbra{\psi_k}{\psi_k})^2\right)},
\end{align}
where the minimization is constrained to all pure state decompositions $\chi=\sum_kp_k\ketbra{\psi_k}{\psi_k}$. If $\conc[\chi]>0$ the Choi state $\chi$ is entangled and the corresponding process has quantum memory. Since no closed analytical expression for the I-concurrence is available, we use here the following negativity-based lower bound $\conc_<[\chi]\leq\conc[\chi]$ given by~\cite{CheAlbFei2005}
\begin{equation}\label{eq:lower_bound_iconcurrence}
\begin{split}
     \conc_<[\chi]=\tilde{N}\mathrm{max}\big\{ &(||\chi^{T_1}||-1), (||\chi^{T_2}||-1)\big\}    
\end{split}
\end{equation}
where $\tilde{N}=\sqrt{\frac{2}{N(N-1)}}$, with $N$ being the dimension of the two parties of the Hilbert space (in our case $N=d^2$), $||\cdot||$ being the trace-norm and $T_{1/2}$ denotes partial transposition with respect to subsystem $1$ or $2$.
Note that any lower bound of the I-concurrence can serve to witness entanglement. Alternative lower bounds have been discussed and compared in the literature, see for example Refs.~\cite{CheAlbFei2005, MinBuc2007, WanFei2025, LuSu2025}. However, in our numerical calculations it turned out that Eq.~\eqref{eq:lower_bound_iconcurrence}
was always the tightest bound.

\section{Details on the map criterion}
\label{sec:mapcriterion_details}
It has been shown in Refs.~\cite{Backer2024Feb,backer2025Entropic} that convex roof constructions can be used to detect a violation of Eq.~\eqref{eq:map_def}, and hence quantum memory.
In this section we will derive a criterion suitable for application to quantum systems beyond qubits, leading to the quantum memory criterion Eq.~\eqref{eq:map_crit_witness}.

{Introducing a general initial system-ancilla state $\rhosa^0$ we can apply a map $\cpt_t$ only to the system part of the bipartite state, analogous to the construction of Choi states}
\begin{align}
\label{eq:bipartite_rho_zero}
    \rho_{\Sys + \Anc}^{(t)} = (\cpt_{t}\otimes \id_\Anc)\rho_{\Sys + \Anc}^0.
\end{align}
In Eq.~\eqref{eq:choi-state} we choose this state to be the maximally entangled Bell state but it can be any entangled state. In fact, in some cases the Bell state may not be the optimal choice here, as shown in Ref.~\cite{backer2025Entropic}.
Following Ref.~\cite{backer2025Entropic}, we define the following functions of a quantum state $\rho$
\begin{align}
\label{eq:F}
    F_f\left[\rhosa\right] = \min_{\{p_k, \rho_k\}} \sum_k p_k f(\rho_k), \notag\\
    F_f^\sharp\left[\rhosa\right] = \max_{\{p_k, \rho_k\}} \sum_k p_k f(\rho_k),
\end{align}
with $f$ being a positive function which is non-increasing under local operations on the system and $\{p_k, \rho_k\}$ describing pure state decompositions of $\rhosa$.

It has been shown in Ref.~\cite{Backer2024Feb} that quantum memory can be detected if one observes
\begin{align}
    \label{eq:qm_witness_maps}
    \eoa\left[\rhosa {(t_1)}\right] < \eof\left[\rhosa {(t_2)}\right].
\end{align}
Choosing the function $f$ to be the concurrence $\conc$ we arrive at an analytically computable criterion for the qubit case, examples are given in Ref.~\cite{Backer2024Feb}.
For systems beyond qubits, however, determining the entanglement properties becomes hard and the criterion Eq.~\eqref{eq:qm_witness_maps} cannot be computed directly.
In Ref.~\cite{backer2025Entropic} an entropic criterion derived from Eq.~\eqref{eq:qm_witness_maps} was proposed, where $f$ is chosen as the entanglement entropy. This measure can be computed efficiently in any dimension, including continuous variable systems. Here, we introduce an alternative criterion which turns out to be stronger in the settings investigated in this article.

In order to give a proof for the criterion we use the I-concurrence introduced earlier in Eq.~\eqref{eq:def_i_concurrence} as our function $f$. As mentioned above, suitable upper and lower bounds for the I-concurrence are available that allow for an efficient estimate.
Noting that Eq.~\eqref{eq:lower_bound_iconcurrence}
corresponds to the first line in Eq.~\eqref{eq:F} if we choose subsystem 1 to be the system and subsystem 2 to be the ancilla
\begin{align}
    \label{eq:criterion_lower_bound}
    F_{\tilde{\conc}}\left[\rhosa\right] = \conc .
\end{align}
We can therefore use  the right-hand side in Eq.~\eqref{eq:lower_bound_iconcurrence} as a lower bound.

For the first line in Eq.~\eqref{eq:F} we will use the bound obtained in Ref.~\cite{LiFeiAlbLiu2009} which reads
\begin{align}
    F_{\tilde{\conc}}^\sharp\left[\rhosa\right] \leq  \min \{\sqrt{2 \left(1 - \tr \rho_\Sys^2\right)},\sqrt{2 \left(1 - \tr \rho_\Anc^2\right)}  \}.
\end{align}
We can now introduce the two-time quantity $F_\Delta(t_1, t_2)$:
\begin{align}
    \label{eq:definition_delta_full}
    F_{\Delta}(t_1, t_2) := \min \{&\sqrt{2 \left(1 - \tr \rho_\Sys^2(t_1)\right)},\sqrt{2 \left(1 - \tr \rho_\Anc^2(t_1)\right)}  \} \notag \\ - \tilde{N}\mathrm{max}\big\{ &||\rho_{\Sys + \Anc}^{T_{\Sys}}(t_2)||-1,||\rho_{\Sys + \Anc}^{T_{\Anc}}(t_2)||-1\big\},
\end{align}
where $\tilde{N}=\sqrt{\frac{2}{N(N-1)}}$, $N$ is the dimension of the two parties of the Hilbert space and $T_{\Sys+\Anc}$ is the partial transposition with respect to Hilbert space $\mathcal{H}_{\Sys+\Anc}$ \cite{CheAlbFei2005}. 
Note again that instead of the lower bound on the I-concurrence given by Ref.~\cite{CheAlbFei2005}, other suitable bounds such as presented in for instance Refs.~\cite{MinBuc2007, WanFei2025, LuSu2025} can be used.
We define
\begin{align}
\label{eq:def_delta}
    \Delta(t_1, t_2) = \max\{0, -F_\Delta(t_1, t_2)\},
\end{align}
which leads us to the following theorem:
\begin{theorem}
\label{th:criterion}
    Let \(\cpt_{t_1}\) and \(\cpt_{t_2}\) be two CPT maps on a system and \(\rho_{\Sys + \Anc}^{0}\) an initial joint state of system and ancilla. Let \(\rho_{\Sys + \Anc}^{(t_1)}\) and \(\rho_{\Sys + \Anc}^{(t_2)}\) be the joint states at times \(t_1\) and \(t_2\) as defined in Eq.~\eqref{eq:bipartite_rho_zero}
    Then, if one observes that
    \begin{align}
    \label{eq:criterion}
        \Delta(t_1, t_2) >0
    \end{align}
    with $\Delta(t_1, t_2)$ as in Eq.~\eqref{eq:def_delta}, the dynamics \(\Dyn=(\cpt_{t_1},\cpt_{t_2})\) is not realizable with classical memory.
\end{theorem}
In the main text we choose \(\rho_{\Sys + \Anc}^{(0)}\) to be the maximally entangled state such that the maximization and minimization in Eq.~\eqref{eq:definition_delta_full} can be performed directly.

\section{Numerical convergence}\label{app:conv}
\begin{figure}[!htb]
    \centering
    \includegraphics[width=0.99\linewidth]{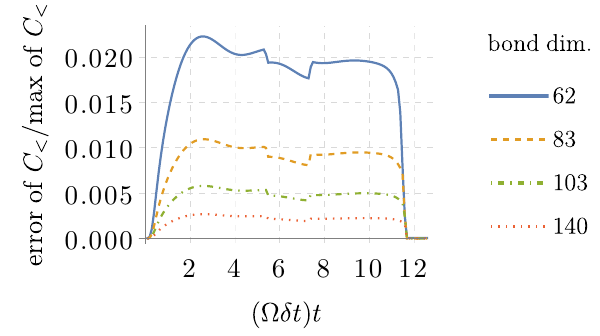}

    \caption{Convergence of $\conc_<$ with respect to the MPO bond dimension of the influence matrix. We considered the dynamics of Fig.~\ref{fig:3}, center panel, $\alpha=0.3$. As a reference we used a calculation with bond dimension 344. The graphs show the absolute value of the deviation in the entanglement criterion with respect to the reference $|\conc_<(t)-\conc_<^\mathrm{ref}(t)|/\mathrm{max}(\conc_<^\mathrm{ref})$.}
    \label{fig:convergence}
\end{figure}

Here we give an example for the convergence of the numerics that was performed using the uniTEMPO algorithm from Ref.~\cite{Link2024May}. We focus on the most difficult case considered in this paper, the stationary process tensor for an Ohmic bath with the strongest coupling $\alpha=0.3$ (Fig.~\ref{fig:3}). Stationary dynamics is more difficult to converge compared with quench dynamics because of additional errors arising from determining the stationary state.
We consider the convergence of the quantum memory criterion with respect to the bond dimension of the MPO influence matrix. The numerical effort scales with the third power of this bond dimension. As can be seen in Fig.~\ref{fig:convergence}, the error for the quantum memory criterion becomes low already for moderate bond dimensions.

\end{document}